# Toward an Affordable Density-Based Measure for the Quality of a Coupled Cluster Calculation

*Published as part of The Journal of Physical Chemistry A special issue "John F. Stanton Memorial Issue".*

Gregory H. Jones,* Kaila E. Weflen, and Jan M. L. Martin*

**ABSTRACT:** We propose two new diagnostics for the degree to which static correlation impacts the quality of a coupled cluster calculation. The first is the change in the Matito static correlation diagnostic $\overline{I_{\mathrm{ND}}}$ between CCSD and CCSD(T), $\Delta\overline{I_{\mathrm{ND}}}[(\mathrm{T})] = \overline{I_{\mathrm{ND}}}[\mathrm{CCSD(T)}] - \overline{I_{\mathrm{ND}}}[\mathrm{CCSD}]$. The second is the ratio of the same and of the corresponding change in the total correlation diagnostic $\overline{I_{\mathrm{T}}} = \overline{I_{\mathrm{ND}}} + \overline{I_{\mathrm{D}}}$, i.e., $r_I[(\mathrm{T})] = \Delta\overline{I_{\mathrm{ND}}}[(\mathrm{T})]/\Delta\overline{I_{\mathrm{T}}}[(\mathrm{T})]$. The first diagnostic can be extended to higher-order improvements in the wave function, e.g., $\Delta\overline{I_{\mathrm{ND}}}[(\mathrm{Q})] = \overline{I_{\mathrm{ND}}}[\mathrm{CCSDT(Q)}] - \overline{I_{\mathrm{ND}}}[\mathrm{CCSDT}]$. In general, a small $\Delta\overline{I_{\mathrm{ND}}}[\mathrm{level}_1]$ value indicates that at this $\mathrm{level}_1$ of theory, the density is converged and any further changes to the energy come from dynamical correlation, while larger $\Delta\overline{I_{\mathrm{ND}}}[\mathrm{level}_2]$ indicates that the density is still not converged at $\mathrm{level}_2$ and some static correlation remains. $r_I[(T)]$ is found to be a moderately good predictor for the importance of post-CCSD(T) correlation effects.

$$\Delta\overline{I_{\mathrm{ND}}}[(T)] = \overline{I_{\mathrm{ND}}}[\mathrm{CCSD(T)}] - \overline{I_{\mathrm{ND}}}[\mathrm{CCSD}]$$

$$\Delta\overline{I_{\mathrm{D}}}[(T)] = \overline{I_{\mathrm{D}}}[\mathrm{CCSD(T)}] - \overline{I_{\mathrm{D}}}[\mathrm{CCSD}]$$

$$r_I[(T)] = \frac{\Delta\overline{I_{\mathrm{ND}}}[(T)]}{\Delta\overline{I_{\mathrm{ND}}}[(T)] + \Delta\overline{I_{\mathrm{D}}}[(T)]}$$

| | Correlation type: dynamical | mildly static | signif. static |
|---|---|---|---|
| $\Delta\overline{I_{\mathrm{ND}}}[(T)]$ | $\lesssim 0.005$ | between | $\gtrsim 0.01$ |
| $r_I[(T)]$ | $\lesssim 0.3$ | between | $\gtrsim 0.5$ |

## 1. INTRODUCTION

Electron correlation, while it accounts for less than one percent of atomic and molecular total energies, has a disproportionately large impact on molecular properties (e.g., between 20 and 180% of small-molecule bond energies). As discussed in, e.g., ref 1., the roughly $\propto \rho^{1/3}$ dependence of the correlation potential on the electron density $\rho$ ensures that the correlation energy increases pronouncedly when atoms are brought together into molecules, or when electrons are added to the system.

The semiroutine evaluation of near-exact correlation energies (and derivative properties) of small molecules has become a reality thanks to patient development of wave function electronic structure approaches, particularly coupled cluster theory.[2] One specific coupled cluster method, CCSD(T),[3,4] offers an unusually felicitous compromise between accuracy and computational cost, and has become known as "the gold standard of quantum chemistry" (T. H. Dunning, Jr.,). However, it is fairly well-known (e.g.,[5−12]) that this good performance is the result of an error compensation between, on the one hand, the neglect of (usually antibonding) higher-order connected triple excitations and, on the other hand, the complete neglect of (universally bonding) connected quadruple excitations.

This compensation breaks down when nondynamical correlation is present. The term was first coined by Sinanoğlu;[13] in the 1990s, the late lamented Björn O. Roos would compress it to "static correlation" during lectures,[14] while the physics literature tends to prefer the term "strongly correlated electrons" (e.g., ref 15.).

The concept is probably best illustrated by comparing four archetypes (see also Hollett and Gill[16] and Martin[1]):

1. the He isoelectronic series: short-range, large HOMO−LUMO gap; the HF determinant is a good zero-order representation of the wave function Ψ; purely *dynamical correlation*;
2. interaction of two He atoms at long distance: long-range version of the above, *dispersion*;
3. the Be isoelectronic series: short-range, small relative HOMO−LUMO gap $(\epsilon_{2p} - \epsilon_{2s})/\epsilon_{2s}$ between 2s and 2p orbitals; excited determinants become prominent; *type B static correlation*;[16]
4. dissociating $H_2$ or $N_2$, or $C_2$ and $O_3$ at equilibrium geometries; longer-range, small HOMO−LUMO gap; a single determinant is a poor zero-order representation of Ψ; *type A static correlation*[16]

The latter is by far the one that causes the most sleepless nights in practical quantum chemists. Hence there have been considerable efforts to develop diagnostics for this problem; for reviews see refs 17,18.

Informally, a number of wave function parameters were used early on for this purpose. For instance, the coefficient of the reference determinant in a single-reference CI calculation, or the sum of squared reference coefficients for a multireference calculation. In coupled cluster theory, intermediate normalization is generally employed, and hence the reference determinant has a unity amplitude by definition. Informally, max$|T_2|$(the largest absolute doubles amplitude) was thus used instead.

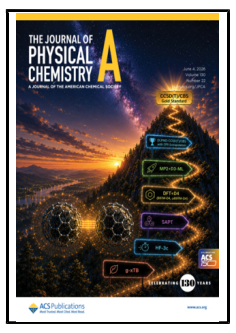

Lee and Taylor[19] proposed the first general-purpose static correlation measure, namely the familiar "$\mathcal{T}_1$ diagnostic":

$$\mathcal{T}_1 = \sqrt{\frac{\mathbf{T}_1^{\mathrm{T}}\mathbf{T}_1}{N_{\mathrm{electrons}}}} \tag{1}$$

I.e., the Euclidean norm of the single excitation amplitudes, divided by the square root of the number of electrons correlated. (The denominator ensures that $\mathcal{T}_1$ for $n$ noninteracting copies of the molecule is independent of $n$, and hence ensures at least approximate size-*in*tensivity.) In follow-up papers,[20,21] the definition of $\mathcal{T}_1$ was extended to open-shell ROHF reference wave functions. A kind of $\mathcal{T}_2$ version has been proposed in ref 22.

Janssen and Nielsen[23] pointed out that $\mathcal{T}_1$ is somewhat vulnerable to "dilution". For example, a reaction center with a lot of static correlation may have an elevated $\mathcal{T}_1$, but attach it to a long aliphatic chain and the $\mathcal{T}_1$ will be quenched. In an attempt to eliminate this problem, they instead proposed to use the matrix 2-norm of the singles amplitudes rectangular matrix $T_1$. (For the outer product of a 1-column vector, i.e., a rank one matrix, the matrix 2-norm and the Frobenius 2-norm — a.k.a., "the Euclidean norm of a matrix" — are equivalent. This is no longer the case when the vector is replaced by a rectangular matrix, such as that of the $T_1$ or $T_2$ amplitudes.)

$$D_1 = \left\|\mathbf{T}_1\right\|_2 = \max_{\|\mathbf{x}\|_2=1}\left\|\mathbf{T}_1\mathbf{x}\right\|_2 = \sqrt{\lambda_{\max}(\mathbf{T}_1\mathbf{T}_1^{\mathrm{T}})} \tag{2}$$

where $\mathbf{x}$ is an arbitrary vector with unit length and $\lambda_{\max}(\mathbf{A})$ is the largest eigenvalue of matrix $\mathbf{A}$. It is easily shown, by considering the example of a complex between BN and $n$-octane at infinite distance, that $D_1$[BN··· $n$-octane]=$D_1$[BN], and hence no "dilution" takes place.

The authors later extended this concept to the doubles amplitudes $T_2$:[24]

$$D_2 = \left\|\mathbf{T}_2\right\|_2 = \max_{\|\mathbf{x}\|_2=1}\left\|\mathbf{T}_2\mathbf{x}\right\|_2 = \sqrt{\lambda_{\max}(\mathbf{T}_2\mathbf{T}_2^{\mathrm{T}})} \tag{3}$$

In the original W4 paper,[11] one of us (JMLM) introduced two pragmatic energy-based diagnostics. One was the percentage of the total atomization energy (TAE) accounted for by parenthetical triple excitations (T):

$$\%\mathrm{TAE}[(\mathrm{T})] = \frac{\mathrm{TAE}[\mathrm{CCSD(T)}] - \mathrm{TAE}[\mathrm{CCSD}]}{\mathrm{TAE}[\mathrm{CCSD(T)}]} \times 100\% \tag{4}$$

The other was the percentage of the atomization energy accounted for at the SCF level, or equivalently, by electron correlation, which is 100%−%TAE[SCF].

$$\%\mathrm{TAE}[\mathrm{corr}] = \frac{\mathrm{TAE}[\mathrm{CCSD(T)}] - \mathrm{TAE}[\mathrm{SCF}]}{\mathrm{TAE}[\mathrm{CCSD(T)}]} \times 100\% \tag{5}$$

$$= 100\% - \%\mathrm{TAE}[\mathrm{SCF}] \tag{6}$$

Fogueri et al.[17] observed that when calculating molecular total atomization energies by hybrid DFT functionals, the dependence of TAE on the percentage of "exact" Hartree–Fock-like exchange is almost perfectly linear, and the slope is clearly linked to the degree of static correlation. They hence proposed using the normalized slope $A$ as a static correlation diagnostic; it was subsequently found[17,25] that this $A$ diagnostic is statistically very similar to %TAE[(T)] and %TAE[corr].

In ref 25. we proposed %TAE$_X$[TPSS@HF] − %TAE$_X$[HF] as a diagnostic — that is, the difference between the exchange contribution to TAE at the TPSS DFT level with Hartree–Fock orbitals and the same evaluated outright at the SCF level. Almost identical information is contained in %TAE$_X$[TPSS] − %TAE$_X$[TPSS@HF], and it is not difficult to grasp why both are statistically very similar to the $A$ diagnostic.

Another family of diagnostics entails the natural orbital (NO) occupations, i.e., the eigenvalues of the 1RDM (first-order reduced density matrix).[26] Already in 1955, Löwdin and Shull[27] proposed their use as measures for correlation strengths.

The von Neumann correlation entropy goes back a long way (see, e.g., ref 28. for an overview). It amounts to

$$S_{\mathrm{corr}} = -\sum_{i,\sigma} n_{i,\sigma} \ln n_{i,\sigma} \tag{7}$$

where the $n_i$ are the natural orbital occupations and the index $\sigma$ runs over spins $\alpha$, $\beta$. An approximately intensive modification we have considered in the past[17,25,29] is $S_{\mathrm{norm}} = S_{\mathrm{corr}}/N_{\mathrm{electrons}}$. We note in passing a recent series of papers[30−34] concerning two-particle correlation entropies derived from the two-particle density matrix, as well as a paper by Evangelista[35] on mutual correlation using density matrices and cumulants.

At the other extreme of simplicity, Truhlar and co-workers proposed[36] the following diagnostic based on the NO occupations of the frontier orbitals:

$$M_{\mathrm{diag}} = \frac{1}{2}\left(2 - n_{\mathrm{HDOMO}} + n_{\mathrm{LUMO}} + \sum_{j\in\mathrm{SOMO}} |n_j - 1|\right) \tag{8}$$

For the special case of a 2-in-2 closed-shell CASSCF, $M = n_{\mathrm{LUMO}}$; the von Neumann correlation entropy reduces to the same for small $n$.

Matito and co-workers[37] applied an expression for the deviation from idempotency of the density matrix to a 2-electron model system, and ultimately arrived at measures for dynamical ($I_{\mathrm{D}}$) and nondynamical ($I_{\mathrm{ND}}$) correlation, as well as total correlation $I_{\mathrm{T}} = I_{\mathrm{ND}} + I_{\mathrm{D}}$. In a later modification[38] they added denominators to ensure approximate size-intensivity, denoted with an overbar:

$$\overline{I_{\mathrm{T}}} = \frac{1}{2N_{\mathrm{electrons}}}\sum_{i,\sigma}[n_{i,\sigma}(1 - n_{i,\sigma})]^{1/2} \tag{9}$$

$$\overline{I_{\mathrm{ND}}} = \frac{1}{N_{\mathrm{electrons}}}\sum_{i,\sigma} n_{i,\sigma}(1 - n_{i,\sigma}) \tag{10}$$

$$\overline{I_{\mathrm{D}}} = \overline{I_{\mathrm{T}}} - \overline{I_{\mathrm{ND}}} \tag{11}$$

Ref 38 also proposed

$$I_{\mathrm{ND}}^{\max} = \max_{i,\sigma}\{n_{i,\sigma}(1 - n_{i,\sigma})\} = n_p(1 - n_p) \tag{12}$$

where $P$ denotes the natural orbital with the maximal contribution to $I_{\mathrm{ND}}$. For obvious reasons, a very high coefficient of determination $R^2$ between $I_{\mathrm{ND}}^{\max}$ and $M_{\mathrm{diag}}$ was found[25,29] to exist. However, it was also found[18,25] that $I_{\mathrm{ND}}^{\max}$ is statistically very similar to $D_2$.

Very recently, Stanton, in his final paper completed during his lifetime, proposed[29] the DAD (density asymmetry diagnostic) based on the deviation from hermiticity of the unrelaxed coupled cluster 1RDM. For full CI (FCI), the exact solution within the given finite basis, the 1RDM is Hermitian (or, for real orbitals, indeed symmetric) and DAD vanishes; this is not the case for truncated CC. Stanton then conjectured a link between how far a given truncated CC is from FCI, and the DAD computed at this level:

$$\mathrm{DAD} = \frac{\left\| D_p^q - D_p^{qT} \right\|_F}{\sqrt{N_{\mathrm{electrons}}}} \tag{13}$$

in which $F$ denotes the Frobenius norm and

$$D_p^q \equiv \langle 0|(1+\Lambda)\exp(-\hat{T})\{p^\dagger q\}\exp(\hat{T})|0\rangle \tag{14}$$

for fully iterative CC methods with truncation of the cluster operator being the only approximation (e.g., CCSD, CCSDT, CCSDTQ, ...). The reader is referred to the literature for the unrelaxed density expressions for CC methods including approximate iterative corrections[39,40] or perturbative corrections.[4,40−43] Statistically (see Table S3 in ref 29) DAD was found to be most similar to $\mathcal{T}_1$ among all the other diagnostics.

The unique feature of DAD is that, unlike other diagnostics, it can act as a gauge for systematic improvement of the coupled cluster level for a given system. However, while a clear link existed within a given system between DAD and remaining error in the correlation energy, energy-based diagnostics were disappointingly dissimilar. (A density-based measure for the closeness of truncated CC solutions to exact solutions was proposed in ref 44., but it obviously presupposes the availability of an (almost) exact solution.)

In the present paper, we shall propose a series of diagnostics based on the changes to the Matito correlation measures between successive levels of theory. These diagnostics not only can be used to gauge convergence with respect to the treatment of nondynamical correlation (and unlike DAD are not specific to CC), but also create a bridge with energy-based diagnostics.

A brief remark is perhaps in order at the end of this introduction. Matito and co-workers[18] distinguish between static correlation measures and diagnostics. Two criteria they impose for the latter are (a) strict size-intensivity and (b) insensitivity to "dilution" by the presence of large moieties with predominantly dynamical correlation effects (such as long alkane chains, multiple water molecules,⋯). By this restrictive definition, only $n_{\mathrm{HOMO}}$, $n_{\mathrm{LUMO}}$, $I_{\mathrm{ND}}^{\max}$, $D_1$, $D_2$, and such would qualify as "diagnostics".

## 2. COMPUTATIONAL METHODS

Most calculations in this paper were carried out using a development version of the CFOUR program system.[45] The required open-shell analytical derivatives were implemented by one of us (GHJ).[46] Some diagnostics such as $D_1$[23] and $D_2$[24] were obtained as byproducts of CCSD calculations using MOLPRO.[47]

The molecules considered are the closed-shell subset of the 200-species W4−17 thermochemical benchmark.[48] These span a range of inorganic and organic molecules, first-row and second-row (including "pseudohypervalent" species in which the 3d acts as an "honorary valence orbital"[50]), and range from essentially purely dynamical correlation (such as $H_2O$ and $SiF_4$) to strong static correlation (such as $O_3$, $S_4$, $C_2$, and BN).

Basis sets considered are the Dunning correlation consistent[51,52] family. Specifically, aside from cc-pVDZ, cc-pVTZ, and cc-pVQZ (correlation consistent polarized double/triple/quadruple-$\zeta$, respectively) we also considered cc-pVDZ with all polarization/angular correlation functions removed, which we denoted cc-pVDZ(p,s).

The approximate CC methods considered include CCSD;[53] CCSD(T);[3,4] full CCSDT;[54,55] CCSDT(Q);[56] and full CCSDTQ.[57]

For all diagnostics based on the NO occupation number or 1RDM, the so-called "relaxed" density, including the contributions of orbital relaxation with respect to an arbitrary perturbation, with the exception of the DAD, where the diagnostic was defined specifically as the "unrelaxed" density corresponding to an asymmetric expectation value (eq 14).

## 3. RESULTS AND DISCUSSION

Full results can be found as a spreadsheet in the Supporting Information (SI).

### 3.1. Closed-Shell Subsets of W4-17 and W4-11

We will focus first on a number of sequences to show trends (Figure 2).

Consider first the sequence $C_2H_6$ to $C_2H_4$ to $C_2H_2$ to $C_2$. The %TAE[(T)] increases along the series, as expected; this is a bit murkier for DAD[CCSD], with $C_2H_6$ and $C_2H_4$ not very different, but quite clear for DAD[CCSDT]. $\mathcal{T}_1$ sort-of shows this trend as well, but $D_1$ and $D_2$ at first sight do not.

Let us now define the correlation measures:

$$\Delta\overline{I_{\mathrm{ND}}}[(T)] = \overline{I_{\mathrm{ND}}}[\mathrm{CCSD(T)}] - \overline{I_{\mathrm{ND}}}[\mathrm{CCSD}] \tag{15}$$

$$\Delta\overline{I_{\mathrm{T}}}[(T)] = \overline{I_{\mathrm{T}}}[\mathrm{CCSD(T)}] - \overline{I_{\mathrm{T}}}[\mathrm{CCSD}] \tag{16}$$

and additionally, the ratio between the (T) increments for nondynamical and total correlation:

$$r_I[(T)] = \frac{\Delta\overline{I_{\mathrm{ND}}}[(T)]}{\Delta\overline{I_{\mathrm{T}}}[(T)]} \tag{17}$$

By analogy, we can define $\Delta\overline{I_{\mathrm{ND}}}[(Q)]$ and $r_I[(Q)]$. (The ratio $\overline{I_{\mathrm{ND}}}/\overline{I_{\mathrm{T}}}$ was considered in ref 18.; the diagnostics defined in eqs 15 and 17 are however new to the present work.)

As a strictly size-intensive and "dilution"-proof diagnostic in the Matito sense, we can also introduce the change in $I_{\mathrm{ND}}^{\max}$ when connected triples are introduced:

$$\Delta I_{\mathrm{ND}}^{\max}[(T)] = I_{\mathrm{ND}}^{\max}[\mathrm{CCSD(T)}] - I_{\mathrm{ND}}^{\max}[\mathrm{CCSD}] \tag{18}$$

One reviewer raised the question of orbital-invariance of the above measures. In short, the measures inherit the orbital-invariance of their parent methods. Given that CCSD(T) and CCSDT(Q) are always calculated using semicanonical orbitals, the investigated measures are not orbital invariant; however, analogs such as $\Delta\overline{I_{\mathrm{ND}}}[\mathrm{CCSDT} - \mathrm{CCSD}]$ would be orbital invariant.

$\Delta\overline{I_{\mathrm{ND}}}[(T)]$ and $r_I[(T)]$ both nicely track the evolution along the "dehydrogenation sequences", as does $\Delta\overline{I_{\mathrm{ND}}}[(Q)]$ at much greater expense. $\Delta I_{\mathrm{ND}}^{\max}[(T)]$, however, like $D_1$ and $D_2$, shows the somewhat counterintuitive result that HCN would have less static correlation than $CH_2NH$, and $C_2H_2$ less than $C_2H_4$. In fact, the same is true for the underlying $I_{\mathrm{ND}}^{\max}$. In all cases this reflects that the relevant NO of the more unsaturated species is degenerate, and by construction all these diagnostics only sample one component. At any rate, the Pearson correlation between $I_{ND}^{max}$ and $\Delta I_{ND}^{max}[(T)]$ is 0.94, compared to 0.95 between $I_{ND}^{max}$ and the largest $T_2$ amplitude, and 0.974 between $I_{ND}^{max}$ and $D_2$. It is hence not obvious that $\Delta I_{ND}^{max}[(T)]$ adds substantial new information over $I_{ND}^{max}$ viz. $D_2$.

The same trend can be seen from propane to propene to propyne ≈ allene to $C_3$. The analogous "dehydrogenation sequences" starting with $N_2H_4$, $CH_3NH_2$, $CH_3OH$ display similar patterns. For the cyclobutane to cyclobutene to cyclobutadiene sequence, the $\mathcal{T}_1$, $D_1$, $D_2$ troika now does show the same trend as the other diagnostics.

For a group of molecules with moderate-to-strong static correlation, like $N_2O$, $O_3$, $S_3$, $S_4$, and the like, $\Delta\overline{I_{ND}}[(T)]$ tracks DAD[CCSDT] better than it does DAD[CCSD]. %TAE[(T)] parts company with $\Delta\overline{I_{ND}}[(T)]$ and DAD for the pseudohypervalent $ClF_5$ molecule; intriguingly, the largest $T_2$ and $T_3$ amplitudes are quite modest, thus siding with $\Delta\overline{I_{ND}}[(T)]$ and DAD for this species.

The original $\overline{I_{ND}}$ predicts some rather puzzling rankings of static correlation: for instance, at the CCSD(T)/cc-pVTZ level, the simple BH diatomic has $\overline{I_{ND}} = 0.090$, which is comparable to $P_2$ and well in excess of clearly multireference species like $S_4$ and $O_3$; the same is observed for AlH. This cannot be ascribed to the normalization and the small number of electrons in these systems: $I_{ND}^{max}$ displays the same behavior, and (consistent with what was said above about the statistical similarity of $I_{ND}^{max}$ and $D_2$) $D_2$ is elevated as well. In contrast, DAD[CCSD] and DAD[CCSDT] for BH and AlH rank among the lowest. $\Delta\overline{I_{ND}}[(T)]$ does not suffer from this problem: its value for BH is comparable to $CH_3PH_2$ and that for AlH comparable to $B_2H_6$, $C_2H_6$, and $CH_3NH_2$, reflecting CCSD's effective treatment of type A static correlation in these particular systems. The top spots are occupied by "the usual suspects" $C_2$, BN, $O_3$, and $S_4$. The same is true, actually, for $\Delta I_{ND}^{max}[(T)]$.

Overall, for the set union of the various sequences considered, $r_I[(T)]$ has a sizable Pearson $R = 0.86$ with %TAE[(T)]. This is quite unlike what was previously found in ref 25, where the energy-based correlation measures were found to be in a different variable cluster from the singles- and doubles-based "diagnostics". Thus, $r_I[(T)]$ bridges between two hitherto disjoint blocks in the $R$ matrix between the variables (see Figure 1).

$\Delta\overline{I_{ND}}[(T)]$ statistically is fairly similar to the DAD variants (especially DAD[CCSDTQ]) and especially to the venerable $\mathcal{T}_1$ "diagnostic". The same is true for $\Delta\overline{I_{ND}}[(Q)]$, which is too expensive for routine evaluation but is statistically almost a factor of 6 smaller than $\Delta\overline{I_{ND}}[(T)]$. Analogous to the progression DAD[CCSD], DAD[CCSDT], and DAD-[CCSDTQ], it demonstrates how the two diagnostic families both taper off toward zero as the level of theory is improved.

How does one need to interpret a small $\Delta\overline{I_{ND}}[(T)]$? (It can be easily verified from the SI that the expression is non-negative for all systems considered.) Effectively, it means that the density does not significantly change between CCSD and CCSD(T), and that the (T) correlation hence primarily reflects dynamical correlation, for which the quasiperturbative (T) treatment should be adequate. Conversely, if $\Delta\overline{I_{ND}}[(T)]$ is significant, then (T) may not be sufficient anymore and an iterative treatment of triples would be indicated.

What happens beyond CCSDT(Q)? We were only able to evaluate full CCSDTQ for a subset of systems (essentially W4-

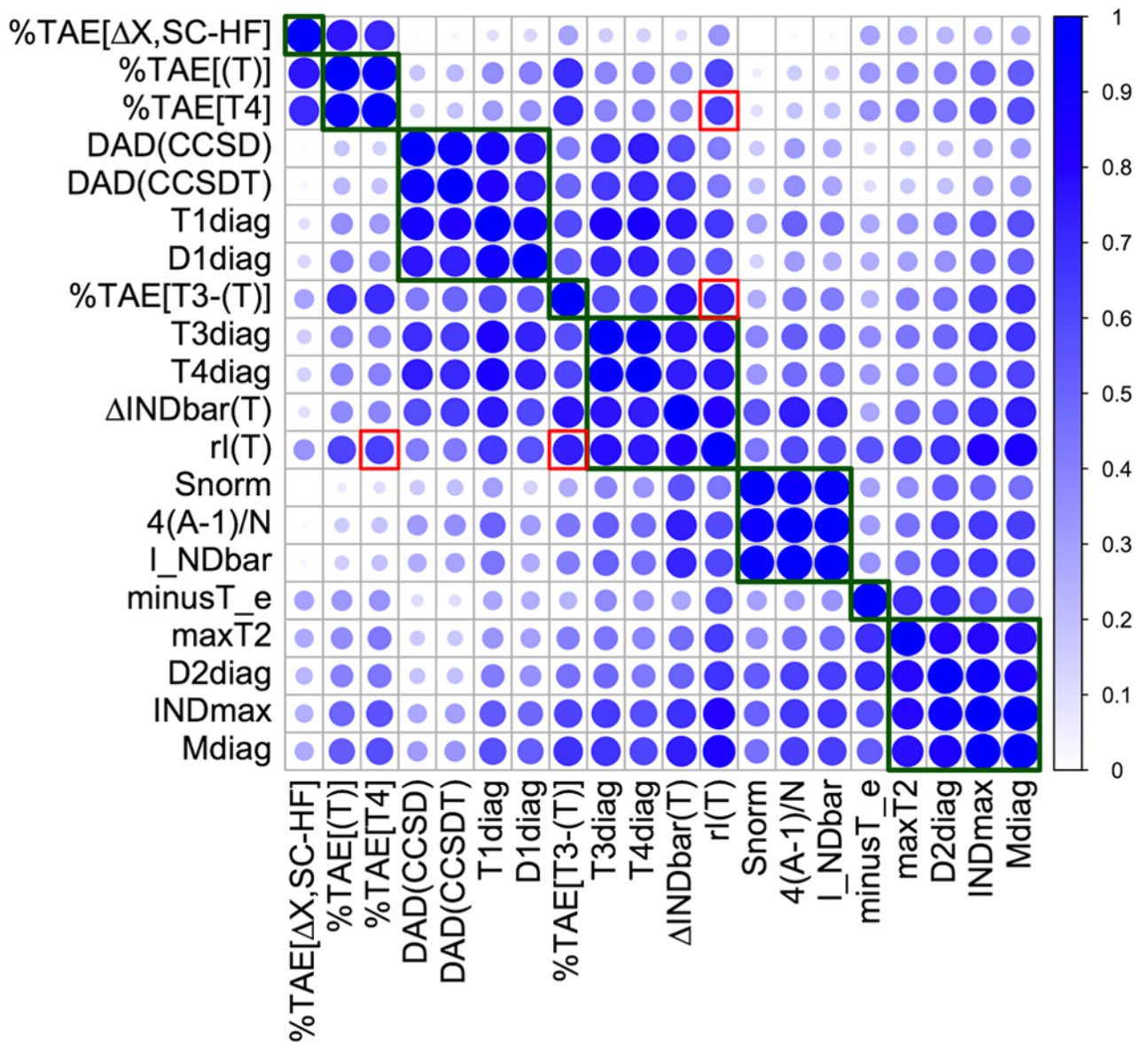

**Figure 1.** Heatmap of $R^2$ values with hierarchical clustering (9 clusters) between different diagnostics for static correlation, on the closed-shell subset of W4-11 with the cc-pVTZ basis set. The diagnostics are, in order: the percentage change in TAE between Hartree−Fock and DFT exchange; percentage of TAE from (T); ditto from connected quadruples; the DAD diagnostic from CCSD and from CCSDT densities; the classic $\mathcal{T}_1$ and $\mathcal{D}_1$ diagnostics; percentage of TAE from higher-order triples; the triples and quadruples analogues of the $\mathcal{T}_1$ singles diagnostic; our newly proposed $\Delta\overline{I_{ND}}[(T)]$ and $r_I[(T)]$ diagnostics; the normalized correlation entropy; (A-1)/N, where $A = \sum T_{ia} + \sum T_{ijab}$, N number of electrons; Matito $\overline{I_{ND}}$; negative the first TD-HF excitation energy; the largest absolute $T_2$ amplitude; the $\mathcal{D}_2$ diagnostic; Matito $\Delta I_{ND}^{max}$; Truhlar $M$ diagnostic.

11[58] plus FNO and ClNO) in the cc-pVDZ basis set. However, with the unpolarized variant of this basis set, cc-pVDZ(p,s), we were able to cover all the closed-shell W4-17 species except for four: $C_2F_6$, $C_2Cl_6$, benzene, and *n*-pentane. In almost all cases, the $\Delta\overline{I_{ND}}[Q-(Q)]$ values are less than 0.0002; the conspicuous exceptions are $C_2$ (−0.0048), BN (−0.0039), and HOClO (+0.0030). Even the strongly multireference $O_3$ and $S_4$ species both render a fairly modest −0.0009. In fact, in such cases as $P_2$, $HClO_3$, $ClF_5$, HCNO, and ClNO, the higher-order quadruples partly compensate for the (Q) contribution.

For the smaller species, we of course can consider the polarized cc-pVDZ basis set. The outliers here are $C_2$ −0.0050, BN −0.0062, $O_3$ −0.0010, $S_4$ −0.0012. Here too, as can be seen in the SI, partial compensation with (Q) takes place. For $HClO_2$, the large $Q-(Q)$ contribution seen with the cc-pVDZ(p,s) basis set is revealed to be an artifact of the small basis set: it is well-known (see, e.g., ref 49 for a review) that for second-row elements in high oxidation states, the 3*d* orbitals acquire some "honorary valence" character as back-bonding recipients from chalcogen or halogen lone pairs. For the diatomics, the corresponding values with the cc-pVTZ basis set are $C_2$ −0.0040, BN −0.0070, SiO −0.0006; once again, just different at the margins from cc-pVDZ.

Do pathological cases like BN still have something going on beyond CCSDTQ? For BN, $C_2$, $N_2$, $P_2$, SiO, and a few more diatomics, we evaluated FCI/cc-pVDZ density matrices using MOLPRO. Even for BN, the effect of higher than quadruple substitutions on $\overline{I_{ND}}$ is just 0.0001, and we can hence assume that nothing truly consequential is happening to the density beyond CCSDTQ.

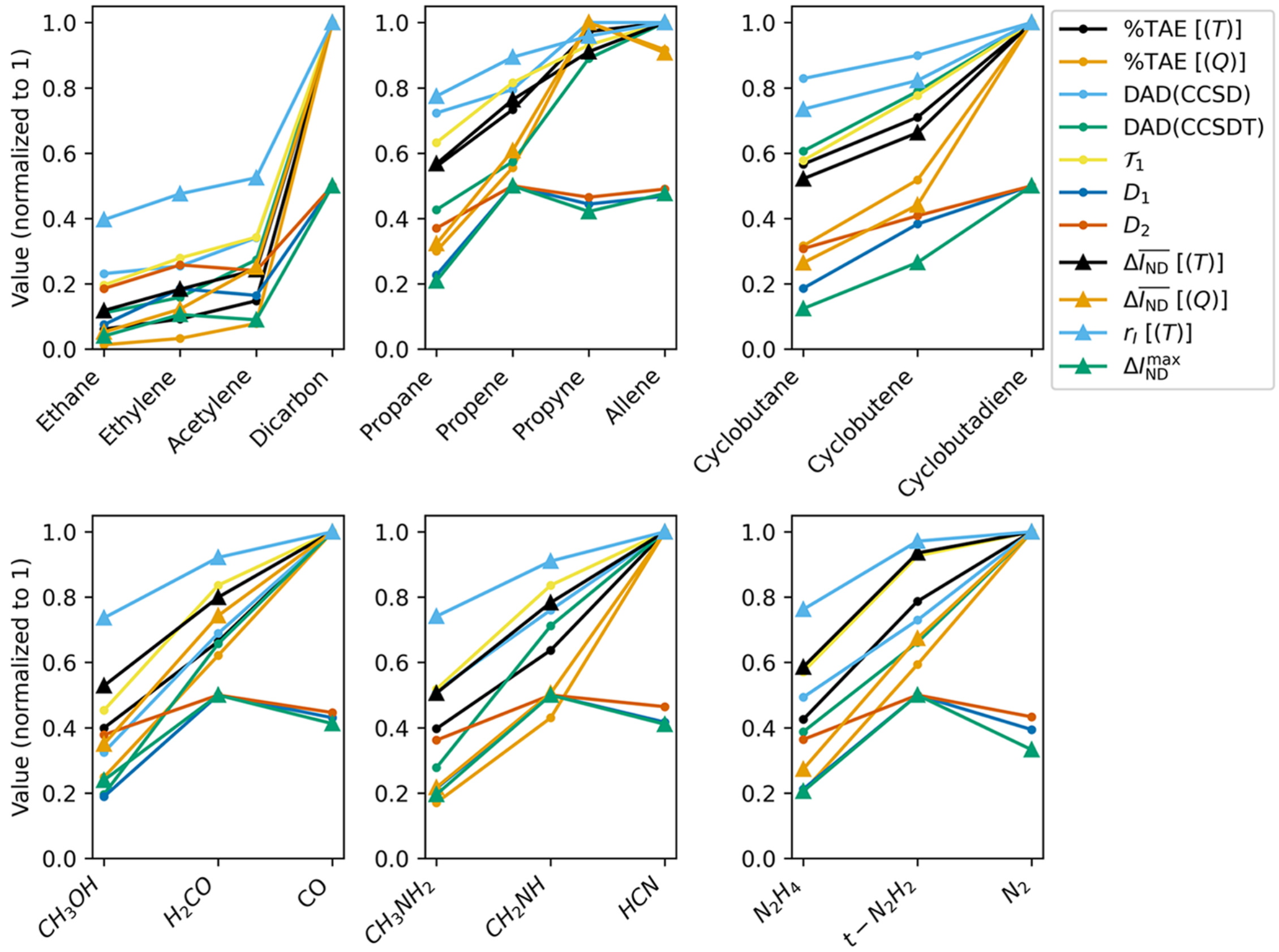

**Figure 2.** Trends in various static correlation measures across various "dehydrogenation sequences". Measures have been normalized to 1 for the largest values, while size-intensive diagnostics have been normalized to 0.5 instead for the sake of graphic separation.

How sensitive are these correlation measures to the basis set? This is illustrated in Figure 3. At the CCSD(T) and even CCSDT levels, cc-pVTZ is a practical option for the entire set. For CCSD(T), so is cc-pVQZ.

$\overline{I_{ND}}$ [CCSD(T)/cc-pVQZ] has an RMS value of 0.0543 over the entire set; the RMS deviation from cc-pVTZ is just 0.0014, indicating that cc-pVTZ can be considered adequately large. The RMS difference with cc-pVDZ is more than five times larger, at 0.0075, while for cc-pVDZ(p,s) 0.0179. $\Delta\overline{I_{ND}}$ [(T)/cc-pVQZ] is 0.0087 RMS over the entire set; again, cc-pVTZ is acceptably close, with an RMSD (root-mean-square difference) of just 0.0006. For cc-pVDZ, this rises to 0.0027, and for cc-pVDZ(p,s) to 0.0043, or nearly half the "signal".

The $r_I$ ratio monotonically decreases with increasing basis set, which makes sense, as the dynamical correlation energy will keep increasing with basis set expansion even past the point where the nondynamical energy is clearly converged. With the cc-pVQZ basis set, the RMS ratio is 0.389, compared to 0.418 for cc-pVTZ, 0.475 for cc-pVDZ, and 0.492 for cc-pVDZ(p,s).

What about higher-order connected triples, i.e., the difference between CCSDT and CCSD(T)? As can be seen in the Supporting Information, not only is their contribution quite small, but it tapers off with increasing basis set. For W4-11 closed, $\Delta\overline{I_{ND}}$ [$T_3$-(T)] decays from 0.00279 RMS for cc-pVDZ(p,s) via 0.00165 for cc-pVDZ and 0.00153 for cc-pVTZ to 0.00150 for cc-pVQZ, but all these values drop down to 0.0005 if $C_2$ and BN are eliminated. In these strongly multireference species, the exaggerated impact of (T) is corrected downward by a strongly *negative* $\Delta\overline{I_{ND}}$ [$T_3$-(T)].

For $\Delta\overline{I_{ND}}$ [(Q)], one would intuitively expect basis set sensitivity to be weaker. Yet, while small, it is nonzero. RMS difference between cc-pVTZ and cc-pVDZ is a measly 0.0004, but that still amounts to one-quarter of the RMS value for $\Delta\overline{I_{ND}}$ [(Q)/cc-pVTZ], 0.0016. Admittedly, the latter is comparable to the likely basis set incompleteness in $\overline{I_{ND}}$ [(T)].

While we are at it, let us consider the basis set sensitivity of the Stanton DAD correlation measure. At the CCSD/cc-pVTZ level, the RMS DAD is 0.01152; the RMS difference with cc-pVQZ is just 0.00047, indicating the cc-pVTZ DAD values are basically converged in terms of the basis set. cc-pVDZ has an RMS difference with cc-pVQZ of 0.00182, which is still good enough for spotting problematic systems; cc-pVDZ(p,s) has an RMS difference of 0.00639, over half the RMS DAD.

At the CCSDT/cc-pVTZ level, the RMS DAD is 0.00373 for the W4-11 subset; the RMS difference (RMSΔ) with cc-pVDZ is 0.00037, clearly still usable, but cc-pVDZ(p,s) with RMSΔ = 0.00176 is not. The partial cc-pVQZ results, with RMSΔ = 0.00013, indicate again that cc-pVTZ DAD is satisfyingly converged with the basis set.

## 3.2. Including Open-Shell Species of W4−11

Almost all of the molecules added to W4-11 in W4-17 are closed-shell, hence we limit ourselves here to W4-11, as additional higher-order correlation contributions such as $T_5$ are available from earlier work.[48]

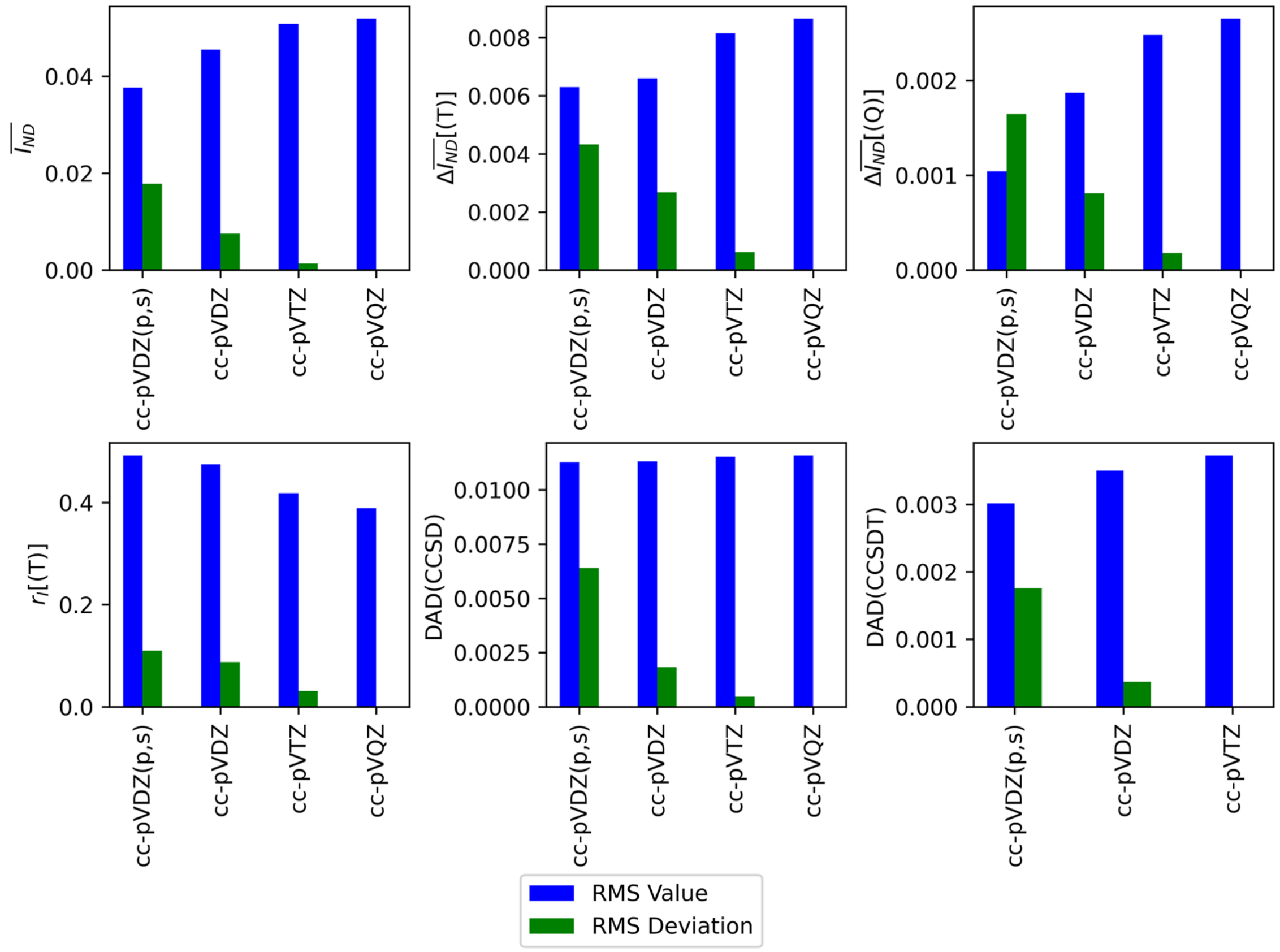

**Figure 3.** Graphical representation of basis set sensitivity of diagnostics for the closed-shell subset of W4-17. The RMS deviation is calculated with respect to the largest basis set in the series. For DAD(CCSDT), the closed-shell subset of W4-11 is shown.

CCSDT(Q) gradients (and hence natural orbitals) in CFOUR are at present only available for closed-shell species, and open-shell extensions for some of the other diagnostics (e.g., $\mathcal{T}_1$) are marred by spin contamination for species such as CN, CCH, $FO_2$, ClOO, and HOOO. Moreover, some of these have multiple UHF solutions, typically one with a lower $\langle S^2\rangle$ and another with a higher one.

In light of the block structure of the determination coefficient matrix in Figure 1, we are limiting ourselves to one representative from certain major blocks (e.g., we are excluding $D_2$ and $M_{diag}$ as they are statistically so similar to $I_{ND}^{max}$). The Pearson correlation coefficients are given in Figure 4 for the cc-pVTZ basis set.

The required post-CCSD(T) contributions were taken from the SI of ref 48.

The strong Pearson $R$ between %TAE[(T)] and %TAE[$T_4$] has been known for some two decades,[11] as has the fact that higher-order triples do not march in lockstep with any one correlation measure. At first sight, the Pearson $R$ between $r_I$[(T)] and %TAE[$T_4$] is somewhat disappointing. If, however, we consider instead the percentage of $T_4$ in *just the correlation part of the total atomization energy*, %TAE$_{corr}$[$T_4$], then the statistical correspondence improves to $R = 0.82$. The second best $R = 0.77$ of %TAE$_{corr}$[$T_4$] is with $I_{ND}^{max}$, itself having a, perhaps surprisingly, high $R = 0.89$ with $r_I$[(T)].

For molecules with significant static correlation, %TAE contributions *beyond* CCSDTQ can exceed 0.5 kcal/mol. Evaluating such connected quintuples and sextuples contributions is computationally truly arduous, with computational scalings of $O^5V^7$ and $O^6V^8$ in terms of the numbers of occupied

| | $r_I$[(T)] | %TAE$_{corr}$[(T)] | %TAE$_{corr}$[$T_4$] | %TAE$_{corr}$[$T_5$+$T_6$] | %TAE[(T)] | %TAE[$T_4$] | DAD | $I_{ND}^{max}$ | $I_{ND}^{bar}$ | %TAE$_{corr}$[$T_3$-(T)] |
|---|---|---|---|---|---|---|---|---|---|---|
| $r_I$[(T)] | 1.00 | 0.60 | 0.66 | 0.58 | 0.55 | 0.56 | 0.39 | 0.84 | 0.61 | 0.24 |
| %TAE$_{corr}$[(T)] | | 1.00 | 0.84 | 0.41 | 0.58 | 0.54 | 0.26 | 0.39 | 0.19 | 0.39 |
| %TAE$_{corr}$[$T_4$] | | | 1.00 | 0.61 | 0.67 | 0.75 | 0.22 | 0.54 | 0.32 | 0.30 |
| %TAE$_{corr}$[$T_5$+$T_6$] | | | | 1.00 | 0.53 | 0.60 | 0.21 | 0.53 | 0.15 | 0.17 |
| %TAE[(T)] | | | | | 1.00 | 0.94 | 0.15 | 0.45 | 0.34 | 0.10 |
| %TAE[$T_4$] | | | | | | 1.00 | 0.22 | 0.52 | 0.19 | 0.10 |
| DAD | | | | | | | 1.00 | 0.33 | 0.21 | 0.18 |
| $I_{ND}^{max}$ | | | | | | | | 1.00 | 0.63 | 0.13 |
| $I_{ND}^{bar}$ | | | | | | | | | 1.00 | 0.12 |
| %TAE$_{corr}$[$T_3$-(T)] | | | | | | | | | | 1.00 |

**Figure 4.** Heatmapped $R^2$ matrix between different diagnostics on the full W4-11 data set (including open-shell species) with the cc-pVTZ basis set.

($O$) and virtual ($V$) orbitals. So can we use any correlation measure to at least predict whether we need to go to this trouble? As it turns out $r_I$[(T)] and %TAE[$T_5$ + $T_6$] have a Pearson $R = 0.76$, just slightly worse than $R = 0.78$ if we used %TAE[$T_4$] instead. Matito $I_{ND}^{max}$ has $R = 0.74$.

Sadly, no single correlation measure lends itself well to predicting $T_3$−($T$); consequently, the same applies to % TAE[post-CCSD(T)] as a whole.

### 3.3. A Test Case: Be Insertion into $H_2$

In the DAD diagnostic paper,[29] Stanton and co-workers considered the classic multireference test case of Be insertion into $H_2$.[59] We are likewise considering it here, using the same cc-pVTZ basis set (which for Be atom was taken from ref 60.). The results are displayed in Figure 5; while reference geometries for points A–J in were taken from ref 59., the figure itself is newly generated from present data. As can be seen there, DAD peaks at E, near the avoided-crossing point, while $r_I[(T)]$ exhibits a still slightly higher value for point F. The latter gets flattened off if we consider $r_I[T_3]$ instead, the profile of which is qualitatively very similar to the correlation energy error $E$[CCSD] − $E$[FCI] — arguably more so than DAD, $M_{\text{diag}}$, or $I_{\text{ND}}^{\text{max}}$.

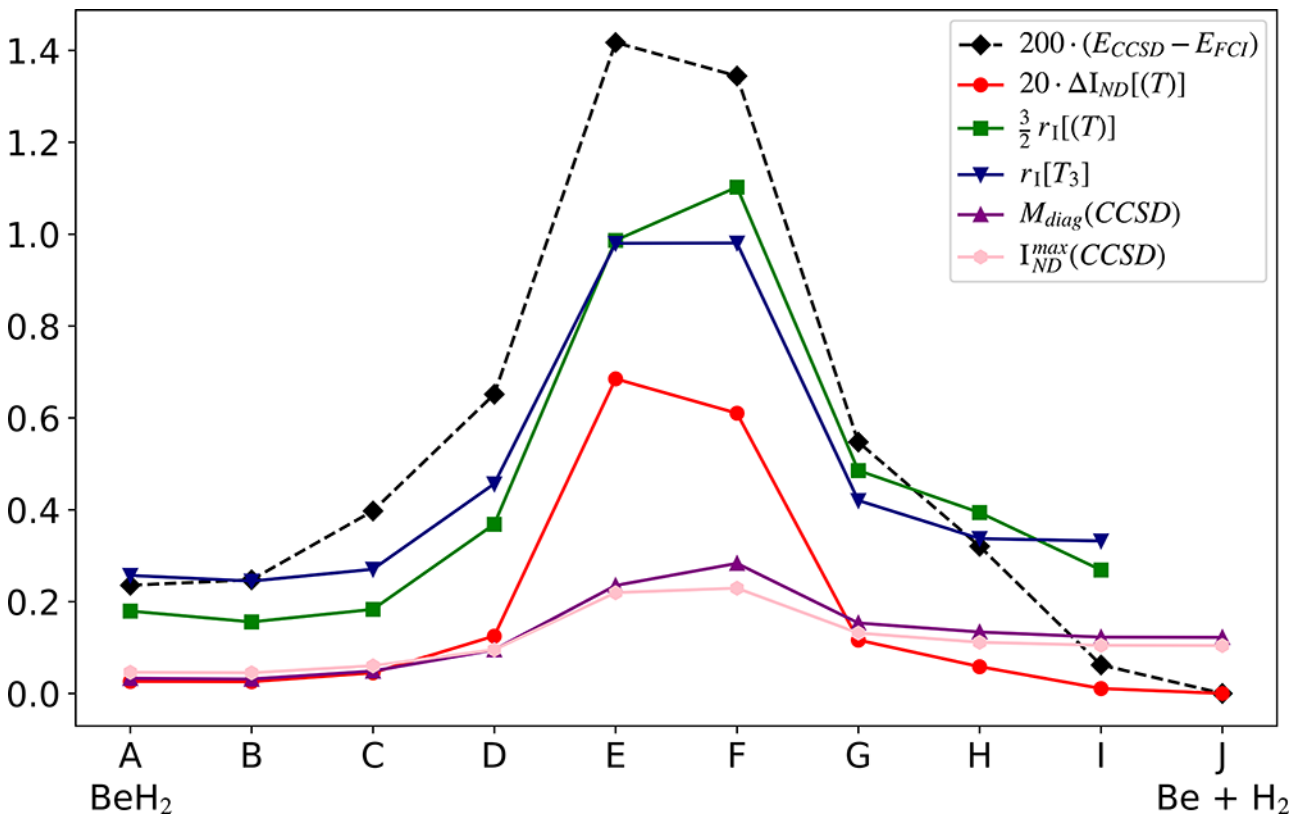

**Figure 5.** Evolution of diagnostics along the reaction profile of Be $+H_2$. Note that several variables have been scaled, as indicated in the legend, to ensure all would be visible in a single plot.

### 3.4. A Model System: $H_4$

A reviewer wondered (citing ref 61) about the behavior of such diagnostics in hydrogen clusters coupled to a singlet and gradually stretched into the dissociation limit, as a model of Mott metal–insulator transitions. In order to address this question, we carried out calculations in the cc-pVTZ basis set on spin-polarized, broken-symmetry $H_4$ chains, at interhydrogen distances of 1.0, 1.25, 1.5, 1.75, 2.0, 2.25, 2.5, and 3.0 Å. Full results can be found in the SI.

Systems like these have very large $T_4$ amplitudes: at $r_{\text{HH}} = 2$ Å, max$|T_4|$ = 0.076, compared to 0.0040 for BN and 0.0014 for $Be_2$. But $H_4$ at this distance also has very large max$|T_1|$= 0.200, max$|T_2|$ = 0.425, and max$|T_3|$= 0.113. (For comparison, the corresponding values for BN are 0.2151, 0.3458, and 0.0309, respectively.)

One feature of the correlation energies of the $H_4$ cluster starkly contrasts with the W4-17 data set. While in the latter, CCSDT(Q) tends to overestimate the importance of quadruples and fully iterative CCSDTQ represents a small(ish) back-correction, CCSDT(Q) in stretched $H_4$ is quite close to CCSDT and higher-order quadruples actually become the lion's share of the $T_4$ contribution. Also, while the Λ coupled cluster[62,63] CCSDT(Q)$_\Lambda$ method is generally quite close to CCSDTQ,[64,65] (see SI), this is not the case for $H_4$ at stretched distances.

In fact, as the cluster is stretched, an ever larger gap opens between $\Delta\overline{I_{\text{ND}}}[(T)]$ from UCCSD and UCCSD(T) calculations (which does not cross the 0.01 threshold until past $r_{\text{HH}}$ = 1.75 Å, while $\Delta\overline{I_{\text{ND}}}[T_3]$ = 0.034 at that distance and 0.037 at 2.0 Å). This behavior is unlike any system in W4-17, where (as we noted above) diagnostics from CCSDT−CCSD—even for troublesome molecules such as BN — are not greatly different from what one obtains from (T).

In contrast, $r_I[(T)]$ and $r_I[T_3]$ at these distances both return values near or even above unity that are bright red flags for strong static correlation.

## 4. CONCLUSION

We propose two new static correlation measures. The first is defined as the change in normalized Matito diagnostic between CCSD and CCSD(T) levels of theory, $\Delta\overline{I_{\text{ND}}}[(T)] = \overline{I_{\text{ND}}}[\text{CCSD(T)}] - \overline{I_{\text{ND}}}[\text{CCSD}]$. It has relatively weak basis set dependence, can be evaluated at fairly low cost compared to the CCSDT(Q) or CCSDTQ calculations that one would like to seek to probe the necessity for, and requires no modification to existing electronic structure codes.

A further extension of the $\Delta\overline{I_{\text{ND}}}$ concept is $\Delta\overline{I_{\text{ND}}}[(Q)] = \overline{I_{\text{ND}}}[\text{CCSDT(Q)}] - \overline{I_{\text{ND}}}[\text{CCSDT}]$. While too expensive for routine evaluation, it can be used to demonstrate tapering off toward zero as the excitation level is increased. Thus, also, in combination with the former, it offers some evidence whether CCSDT(Q) is acceptably close to the exact solution in the basis set at hand.

The second measure is the ratio $r_I[(T)] = \Delta\overline{I_{\text{ND}}}[(T)]/\Delta I_T[(T)]$, where $I_T$ is the total correlation diagnostic of Matito. Unlike the two others, it sidesteps the normalization issue. Moreover, it appears to correlate moderately well with the importance of post-CCSD(T) correlation effects.

Basis set convergence of these diagnostics is moderately rapid, with cc-pVDZ values being usable, and cc-pVTZ values effectively converged. In small basis sets, $r_I$ is overestimated while $\Delta\overline{I_{\text{ND}}}$ is underestimated.

What are safe ranges for $\Delta\overline{I_{\text{ND}}}$ and $r_I[(T)] = \Delta\overline{I_{\text{ND}}}[(T)]/\Delta\overline{I_{\text{T}}}[(T)]$? Analysis of our data reveals that typical alkane and other singly bonded species, dominated by dynamical correlation, will have $\Delta\overline{I_{\text{ND}}}[(T)]$ around 0.004, increasing to 0.006 if there is a double bond (which introduces a bit of static correlation), 0.007–8 if there is a triple bond or cumulenic pair of double bonds, and values in the 0.01 or higher range indicate more significant static correlation. In contrast, $r_I[(T)]$ will be in the 0.3 range for mostly dynamical correlation species, while 0.5 or higher indicates substantial static correlation.

Matito and co-workers[18] (see also Chan[66]) made the argument that there is a fundamental difference between strong correlation *measures* and static correlation *diagnostics*. (This dichotomy is easily illustrated by comparing and contrasting the problematic $Be_2$ molecule with $N^{+++}$ − the latter has a very high $D_2$ diagnostic from type B static correlation between 2*s* and 2*p* orbitals, but CCSD can handle it perfectly well.) They also argue that pragmatic energy-based measures are intrinsically unsuitable as diagnostics. Be that as it may, but for practicing quantum chemists it is obviously crucial to be able to predict, for a given problem, whether CCSD(T) is still the gold standard or whether its coin is being debased by static correlation. In this regard, the interesting feature of $r_I[(T)]$ lies in its reconciling, at fairly modest computational cost, the conflicting goals of measures and diagnostics.

Concerning diagnostics for excited states, we have initial evidence (K. E. Weflen, G. H. Jones, and J. M. L. Martin, to be published) that DAD[EOM-CCSD] and DAD[EOM-CCSDT] can be readily proposed as excited-state extensions of the DAD diagnostic, while the extensions of $\Delta I_{ND}$ will be more problematic.[67]

## ■ ASSOCIATED CONTENT

### Supporting Information

The Supporting Information is available free of charge at https://pubs.acs.org/doi/10.1021/acs.jpca.6c00684.

All relevant computed diagnostics and Python post-processing script for extraction of Matito diagnostics from CFOUR output (ZIP)

## ■ AUTHOR INFORMATION

### Corresponding Authors

**Gregory H. Jones** − *Quantum Theory Project, Department of Chemistry, University of Florida, Gainesville, Florida 32611, United States;* orcid.org/0000-0003-3275-1661; Email: gh.jones@ufl.edu

[§]**Jan M. L. Martin** − *Department of Molecular Chemistry and Materials Science, Weizmann Institute of Science, 7610001 Reḥovot, Israel*; Email: gershom@weizmann.ac.il

### Author

**Kaila E. Weflen** − *Department of Molecular Chemistry and Materials Science, Weizmann Institute of Science, 7610001 Reḥovot, Israel;* orcid.org/0009-0003-2483-8692

Complete contact information is available at: https://pubs.acs.org/10.1021/acs.jpca.6c00684

### Author Contributions

J.M.L.M.: conceptualization, methodology (lead), data curation, analysis (lead), draft (lead); G.H.J.: coding in CFOUR, methodology (supporting), analysis (supporting), draft (supporting); K.E.W.: coding in CFOUR, draft (supporting). All authors contributed to writing—review and editing.

### Notes

The authors declare no competing financial interest.
[§]On sabbatical at: Quantum Theory Project, Department of Chemistry, University of Florida, Gainesville, Florida 32611, United States.

## ■ ACKNOWLEDGMENTS

J.M.L.M. would like to acknowledge inspiring discussions at the Quantum Theory Project, U. of Florida, with its late lamented director, Prof. John F. Stanton (1961-2025). G.H.J. was funded by the National Science Foundation (grant CHE-2430408, "Advances in Coupled Cluster Theory"; deceased PI: John F. Stanton; current PI: Alberto Perez). Research at Weizmann was supported in part by a grant to the Martin group from the Minerva Foundation, Munich, Germany, and by a generous allocation of computer time on the Faculty of Chemistry's HPC facility CHEMFARM, which is supported in part by the Ben May Center for Chemical Theory and Computation (Weizmann Institute of Science).

## ■ REFERENCES